\documentstyle[multicol,prb,aps,epsf]{revtex}
\begin{document}
\title{Electron self-energy in A$_3$C$_{60}$ (A=K, Rb):
Effects of $t_{1u}$ plasmon in GW approximation}

\author  {O. Gunnarsson} 
\address{Max-Planck-Institut f\"ur Festk\"orperforschung,
 D-70506 Stuttgart, Germany}

 \date{\today}

\maketitle

\begin{abstract}
The electron self-energy of the $t_{1u}$ states in A$_3$C$_{60}$ (A=K, Rb) 
is calculated using the so-called GW approximation. 
The calculation is performed    within a model which considers the
$t_{1u}$ charge carrier plasmon at 0.5 eV and takes into account 
scattering of the electrons within the $t_{1u}$ band. A moderate
reduction (35 $\%$) of the $t_{1u}$ band width is obtained. 
\end{abstract}

\begin{multicols}{2}
\section{Introduction}
The alkali doped fullerenes A$_3$C$_{60}$ (A=K, Rb) have a low-energy
charge carrier plasmon at an energy of about 0.5 
eV,\cite{Sohmen,Eyert,Liechtenstein} which essentially results 
from the oscillation of the alkali electrons donated to the $t_{1u}$ band.
This plasmon has a rather strong coupling to the electrons and it is 
believed to play an important role for the anomolously large band width
seen in angular integrated photoemission for these systems,\cite{Knupfer} 
since it can cause large binding energy plasmon satellites. 

It is interesting to calculate the electron self-energy due to the
coupling to the plasmons, to more generally study the effects of 
the plasmons on the electronic properties. We here use the so-called GW 
approximation, where the self-energy is calculated to lowest order
in the screened Coulomb interaction.\cite{Hedin}
This approximation has been widely applied to the electron gas,    
free-electron like system and semiconductors. For these systems
it is found that the self-energy has a moderate effect on the
electron dispersion and that it can somewhat increase or reduce the
effective mass, depending on the system.
A$_3$C$_{60}$, however, has a quite different band structure from
these systems, and it is interesting to ask if the effective mass
and the width of the band might be more strongly modified in this
system.

This issue is, for instance, raised by the results for the 
optical conductivity. Experimentally, the Drude peak is very
narrow,\cite{DeGiorgi} perhaps an order of magnitude narrower
than predicted by a one-particle theory.\cite{Brink}
This may suggest that the effective mass is greatly enhanced 
in A$_3$C$_{60}$. On the other hand, the specific heat has been 
estimated to be small,\cite{Meingast} perhaps even smaller than would
have been expected from band structure calculations,\cite{RevModPhys}
suggesting that the effective mass is not strongly renormalized,
and perhaps even somewhat reduced. The coupling to phonons with the coupling
constant $\lambda_{ph}$ reduces the dispersion by a factor of
 $1/(1+\lambda_{ph})$.
If this  result is directly taken over for the plasmons, we may estimate 
a coupling constant $\lambda_{pl}\sim 2.5$ and a substantial reduction
(more than a factor of three) of the dispersion.  

The coupling to a plasmon studied here corresponds to the GW method
in the plasmon pole approximation, which was first introduced and extensively 
studied by Lundqvist {\it et al.}\cite{Lundqvist} and later by
Overhauser\cite{Overhauser} for the electron gas.
Shirley and Louie have applied the GW approximation to solid
C$_{60}$.\cite{Louie} Since they considered undoped C$_{60}$, the $t_{1u}$
plasmon considered here did not enter their calculation, while the 
contribution to the self-energy in their calculation (a plasmon at 
about 25 eV) is  not included in the model used below. The two calculations
are therefore complementary and address different parts of the physics
of $A_3$C$_{60}$.

The coupling to the plasmon leads to a so-called plasmaron satellite
in the spectral function, where a plasmon has been excited.\cite{Lundqvist}
This satellite is obtained in the GW approximation,\cite{Lundqvist} 
but a more accurate treatment would also
give higher satellites with several plasmons excited.\cite{Langreth,HedinPS}
These higher plasmon satellites are believed to be important for the broad
spectrum seen in angular integrated photoemission for 
A$_3$C$_{60}$.\cite{Knupfer,Migdal} 
The GW approximation can therefore not be expected to give an accurate
spectrum in the satellite region.
We also observe that the Coulomb interaction $U$ between two electrons
on the same C$_{60}$ molecule is large compared with the $t_{1u}$ band 
width $W$,\cite{Sawatzky} which suggests that many-body effects beyond
the GW approximation may be important. For instance, a proper treatment
of the three-fold degeneracy of the $t_{1u}$ orbital is important 
for understanding why these systems are not Mott-Hubbard insulators,
in spite of the large value of $U/W$.\cite{C60Mott}
The GW results for A$_3$C$_{60}$ should therefore be treated with 
a certain caution, but it is still interesting to see if a strong
renormalization of the band width is obtained.

In Sec. II we present the formalism and in Sec. III the model used in
these calculations. The results are presented in Sec. IV and discussed
in Sec. V.

\section{Formalism}

We write the dielectric function of A$_3$C$_{60}$ as
\begin{equation}\label{eq:1}
\epsilon({\bf q},\omega)=\epsilon_0-{\omega_{0p}^2\over m^{\ast}\omega^2},
\end{equation}
where $\epsilon_0$ is the contribution to $\epsilon$ due to all excitations
in the system except the ones inside the $t_{1u}$ band. Since most of these 
excitations have a rather high energy relative to the energy scale of 
interest here (the $t_{1u}$ band width), we assume $\epsilon_0$ to
be energy independent. The second term describes the excitations 
inside the $t_{1u}$ band. 
$\omega_{0p}$ is the plasmon frequency one would 
deduce for free electrons with the same density as the $t_{1u}$ 
electrons and $m^{\ast}$ is 
the band mass of the $t_{1u}$ electrons. The value of $\omega$ where
$\epsilon({\bf q},\omega)=0$ gives the plasmon frequency
\begin{equation}\label{eq:2}
\omega_p={\omega_{0p}\over \sqrt{m^{\ast}\epsilon_0}}.
\end{equation}
Above we have neglected local-field effects as well as the 
${\bf q}$-dependence of $\epsilon_0$. Neither approximation is quite
justified.\cite{Eyert} 
Thus the ${\bf q}$-dependence of $\epsilon_0$
tends to give $\omega_p$ a positive dispersion, while the local-field
effects tend to give a negative dispersion. As far as the plasmon 
frequency is concerned, however, these two effects essentially 
cancel, in agreement with the experimental observation that      
$\omega_p$ has a negligible dispersion.\cite{Eyert}
Eqs.~(\ref{eq:1},\ref{eq:2}) therefore give a good description of the plasmon
frequency, but neglect the substantial broadening of the 
plasmon.\cite{Liechtenstein}

For K$_3$C$_{60}$ the $t_{1u}$ electron density corresponds to the 
electron gas density parameter $r_s=7.3 a_0$ and $\omega_{0p}=2.4$
eV. In an electron gas of this density, the occupied part of the
band is about 0.9 eV, while the calculated full band width of  K$_3$C$_{60}$ 
is about 0.6 eV.\cite{RevModPhys} This corresponds to 
$m^{\ast}\sim 0.9/(0.6/2)= 3$. 
At ${\bf q}=0$ $\epsilon_0\sim4.4$.\cite{eps}  We then find that
$\omega_p\sim 0.66$ eV. This is somewhat larger than the experimental
result, $\omega_p\sim 0.5$ eV,\cite{Eyert} which will be used in the 
following.

The self-energy is in the GW approximation written as\cite{Hedin}
\begin{eqnarray}\label{eq:3}
&&\Sigma_{nn^{'}}({\bf k},\omega)=   \\
&&i\sum_{{\bf k}^{'}n^{''}}\int
{d\omega^{'}\over 2 \pi} {U_{nn^{''},n^{'}n^{''}}({\bf k-k}^{'})
\over \epsilon(\omega^{'})}{e^{i\omega^{'}0^{+}}\over
\omega+\omega^{'}-\varepsilon_{{\bf k}^{'}n^{''}}+\mu_0}, \nonumber
\end{eqnarray}
where $U_{nn^{''},n^{'}n^{''}}({\bf k-k}^{'})$ is the Coulomb matrix      
element between the Bloch states $|{\bf k}n\rangle$ and $|{\bf k}^{'}n^{''}
\rangle$ with the argument ${\bf r}$ and $|{\bf k}n^{'}\rangle$ and 
$|{\bf k}^{'}n^{''}\rangle$ with the argument ${\bf r}^{'}$. 
Here ${\bf k}$ is a wave vector and $n$ a band index.
$0^{+}$ is an positive infinitesimal number, which assures the proper
behavior for large $\omega^{'}$.    $\varepsilon_{{\bf k}^{'}n^{''}}$
is the noninteracting energy of the state $|{\bf k}^{'}n^{''}\rangle$, 
and $\mu_0$ is an average of the self-energy over the Fermi surface.
We split the dynamically screened Coulomb interaction $U/\epsilon$ in two parts
\begin{equation}\label{eq:4}
{U\over \epsilon(\omega)}={U\over \epsilon_0}+{\omega_p U\over 2\epsilon_0}
{2\omega_p \over \omega^2 -\omega_p^2}.
\end{equation}
The first term then gives a statically screened exchange contribution 
to the self-energy. 
\begin{equation}\label{eq:4a}
\Sigma^{x}_{nn^{'}}({\bf k},\omega)=-{1\over \epsilon_0}
\sum_{{\bf k}^{'}n^{''}}^{{\rm occ}} U_{nn^{''},n^{'}n^{''}}({\bf k-k}^{'}),
\end{equation}
while the second part gives a correlation 
contribution. We can interpret $2\omega_p/(\omega^2-\omega_p^2)$ in 
Eq.~(\ref{eq:4}) as a boson Green's function, and 
\begin{equation}\label{eq:5}
g_{nn^{''},n^{'}n^{''}}^2({\bf q})=
{\omega_p\over 2\epsilon_0}U_{nn^{''},n^{'}n^{''}}({\bf q}),
\end{equation}
as a coupling constant, in analogy with previous work.\cite{Lundqvist}
Closing the integration contour in the upper half of the complex
$\omega^{'}$ plane, we obtain the correlation contribution to the 
self-energy. 
\begin{eqnarray}\label{eq:6}
\Sigma^{c}_{nn^{'}}({\bf k},\omega)&&={\omega_p\over 2\epsilon_0}
\sum_{{\bf k}^{'}n^{''}}^{{\rm occ}} {U_{nn^{''},n^{'}n^{''}}({\bf k-k}^{'})
\over \omega-\varepsilon_{{\bf k}^{'}n^{''}}+\mu_0+\omega_p} \\
&&+{\omega_p\over 2\epsilon_0}
\sum_{{\bf k}^{'}n^{''}}^{{\rm unocc}} {U_{nn^{''},n^{'}n^{''}}({\bf k-k}^{'})
\over \omega-\varepsilon_{{\bf k}^{'}n^{''}}+\mu_0-\omega_p} \nonumber
\end{eqnarray}
  
\section{Model}

We consider a model with three $t_{1u}$ orbitals. The hopping matrix
elements between these orbitals have been described in a tight-binding
parametrization,\cite{Orientation,Satpathy} which is used here.
This parametrization is used to calculate the noninteracting band structure
energies $\varepsilon_{{\bf k}n}$ and wave functions
\begin{equation}\label{eq:7}
\psi_{{\bf k}n}({\bf r})=\sum_{\nu=1}^3c^{(n)}_{\nu}\phi_{{\bf k}\nu}({\bf r}),
\end{equation}
where 
\begin{equation}\label{eq:8}
\phi_{{\bf k}\nu}({\bf r})={1\over \sqrt{N}}\sum_{j=1}^N
 e^{i{\bf k}\cdot{\bf R}_j} \Phi_{\nu}({\bf r-R}_j)
\end{equation}
is a Bloch state of a $t_{1u}$ molecular orbital $\Phi_{\nu}({\bf r})$.
There are $N$ molecules with the coordinates ${\bf R}_j$

We further have to specify the matrix elements of the Coulomb 
interaction. We assume
\begin{eqnarray}\label{eq:9}
&&\int d^3 r\int  d^3 r^{'} \Phi_{\nu_1}({\bf r-\bf R})
 \Phi_{\nu_2}({\bf r}-{\bf R})
{e^2\over|{\bf r}-{\bf r}^{'}| }
 \Phi_{\nu_3 }({\bf r}^{'}-{\bf R}^{'}) \nonumber \\
&& \times \Phi_{\nu_4  }({\bf r}^{'}-{\bf R}^{'})
=\delta_{\nu_1\nu_2}\delta_{\nu_3\nu_4}\times\cases{ {e^2\over 
|{\bf R}-{\bf R}^{'}|}, & if ${\bf R} \ne {\bf R}^{'}$;
\cr U_0, & if ${\bf R} =   {\bf R}^{'}$\cr}                 
\end{eqnarray}
For the Coulomb matrix elements between Bloch states we then find
\begin{eqnarray}\label{eq:10}
&&\langle{\bf k}\nu_1{\bf k}^{'}\nu_2|{e^2\over |{\bf r-r}^{'}|}
|{\bf k}\nu_3{\bf k}^{'}\nu_4\rangle \nonumber \\
&&={1\over N}(U_0+\sum_{{\bf R}\ne 0}
{e^{i({\bf k-k}^{'})\cdot{\bf R}}\over |{\bf R}|}) 
\delta_{\nu_1\nu_2}\delta_{\nu_3\nu_4}.
\end{eqnarray}
We replace the sum over ${\bf R}$ by an integral over all space outside
a Wigner-Seitz sphere with the radius $R_0=5.56$ \AA. This gives 
\begin{eqnarray}\label{eq:11}
&&\langle{\bf k}\nu_1{\bf k}^{'}\nu_2|{e^2\over |{\bf r-r}^{'}|}
|{\bf k}\nu_3{\bf k}^{'}\nu_4\rangle  \\
&&={1\over N}\lbrack U_0+{4\pi e^2\over 
\Omega|{\bf k-k}^{'}|^2}{\rm cos} (R_0|{\bf k-k}^{'}|)\rbrack
\delta_{\nu_1\nu_2}\delta_{\nu_3\nu_4},\nonumber
\end{eqnarray}
where $\Omega$ is the volume of the Wigner-Seitz cell.
We have put  $U_0=4$ eV, using a simple estimate based on the radius
of the C$_{60}$ molecule. Alternatively,
we can extend the integration over $R$ over all space, putting $R_0=0$.
In this case we should put $U_0=0$ to avoid double counting.

\noindent
\begin{figure}[bt]
\unitlength1cm
\begin{minipage}[t]{8.5cm}
\centerline{\epsfxsize=3.375in \epsffile{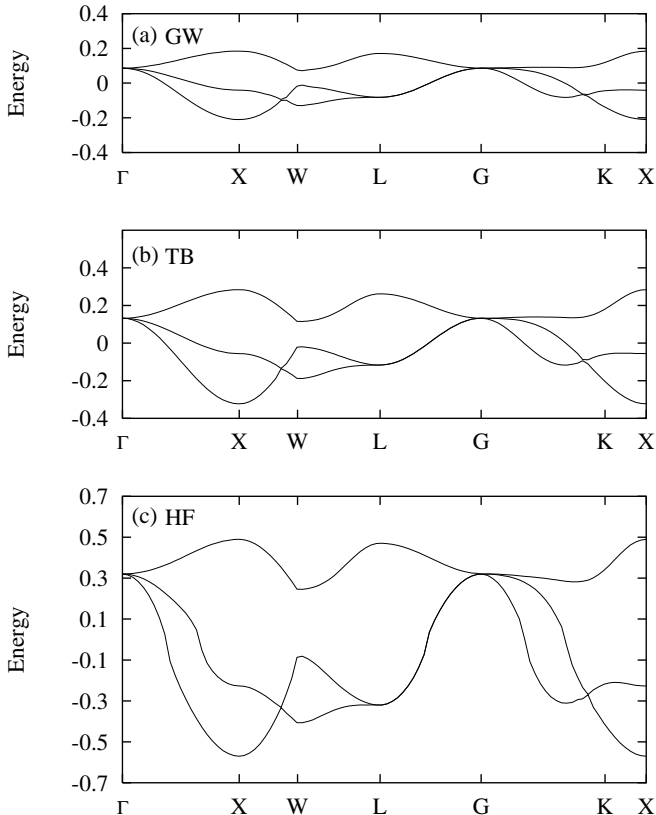}}
\caption[]{\label{fig1}
Quasi-particle (GW) (a) band structure for the $t_{1u}$ band 
compared with the noninteracting tight-binding (TB) (b) and Hartree-Fock
(HF) (c) band structures. Relative to the TB band structure,
the GW band width is reduced by about 35 $\%$, 
 while the HF width is increased by about 75 $\%$.} 
\end{minipage}
\hfill
\end{figure}

\section{Results}
We have calculated the self-energy as described above. The quasi-particle
energy $E_{{\bf k}n}$ is obtained by solving the equation       
\begin{equation}\label{eq:12}
E_{{\bf k}n}=\varepsilon_{{\bf k}n}+\Sigma({\bf k},E_{{\bf k}n}),
\end{equation}
where $\varepsilon_{{\bf k}n}$ is the band structure energy.
We can also obtain the quasi-particle weight as
\begin{equation}\label{eq:13}
Z_{{\bf k}n}={1\over 1-\delta\Sigma({\bf k},\omega)/\delta \omega},
\end{equation}
where the derivative is evaluated at the quasi-particle energy.
The resulting quasi-particle (GW) band structure is compared with 
the noninteracting tight-binding (TB) and Hartree-Fock (HF)
band structures in Fig. \ref{fig1}.
The figure shows a moderate reduction ($\sim 35 \%$) of the GW 
band width relative to the TB  width, while the HF band width is almost
a factor of two larger ($75 \%$) than the TB result.                          

The quasi-particle strength is relatively small, $Z\sim 0.4-0.5$. This is 
smaller than what is found in electron gas calculations at metallic 
densities. Since the three $t_{1u}$ electrons correspond to a very low    
density with $r_s\sim 7$, this small value of $Z$ is, however, 
not very surprising. Actually extrapolation of earlier\cite{Hedin} GW 
calculations for the electron gas to $r_s=7.3$ suggest $Z\sim 0.52$.
The small value of $Z$ implies that a substantial 
spectral weight is transferred to satellites.
 
Replacing $U_0=4$ eV and $R_0=5.56$ \AA \ in the model (\ref{eq:11}) by 
$U_0=0$ and $R_0=0$, e.g., a pure $1/q^2$ interaction gives a small 
change of the quasi-particle band structure. This suggests   that the
results are not very sensitve to the details of the model.

\section{Discussion}

The results in Fig. \ref{fig1} illustrate that there is a large 
cancellation between exchange and correlation effect. This is also
observed in the electron gas, although in that case the cancellation
is more complete than here.\cite{Hedin,Lundqvist}

Of particular interest is the energy dependence of the self-energy.
For the sake of the discussion, we neglect the ${\bf q}$-dependence 
of the coupling constant $g({\bf q})$ (Eq.~(\ref{eq:5}))
and replace it by its average over the Brillouin zone. We further assume 
that the plasmons
couple to a nondegenerate band of width $2B$ and with a constant density
of states.
The correlation part of the self-energy is then given by 
(Eqs.~(\ref{eq:6}))
\begin{equation}\label{eq:14}
{\rm Re} \Sigma^c_0(\omega)={\lambda \omega \over 2}\lbrack 
{\rm ln}|{\omega+B+\omega_p\over \omega+\omega_p}|+
{\rm ln}|{\omega-\omega_p \over \omega-B-\omega_p}\rbrack
\end{equation}
where 
\begin{equation}\label{eq:15}
\lambda={2g^2\over \omega_p}N(0)={g^2\over \omega_p B}
\end{equation}
is a coupling constant defined as for the electron-phonon coupling and
$N(0)=1/(2B)$ is the density of states. 
In the present case we have $\lambda\sim 2.5$.
In the limit $|\omega| <<\omega_p<<B$
we have
\begin{equation}\label{eq:16}
{\rm Re}\Sigma^c_0(\omega)=-\lambda \omega, 
\hskip 0.5cm |\omega|<<\omega_p<<B.
\end{equation}
In the opposite limit where $\omega_p>>B, |\omega|$ we have
\begin{equation}\label{eq:17}
{\rm Re}\Sigma^c_0(\omega)=-({g\over \omega_p})^2\omega
\hskip 0.5cm |\omega|, B<<\omega_p.
\end{equation}
The coupling constant $(g/\omega_p)^2\sim 1.4$ in the present case.
The ``phonon''-like formula in Eq.~(\ref{eq:16}) predicts that the 
quasi-particle weight is $Z\sim 0.29$ while the formula (\ref{eq:17})
predicts $Z\sim 0.42$. The latter result fits the actual calculations
nicely, not too surprisingly in view of the parameter range
considered here ($\omega_p=0.5$ eV, $B=0.3$ eV).
The fact that the system is not in the ``phonon'' parameter range
($\omega_p<<B$) therefore means that correlation effects lead to a 
less drastic narrowing of the band. This narrowing is furthermore 
to a substantial part compensated by exchange effects.

Nevertheless, the narrowing of the band is larger than found in the
GW approximation for the electron gas at metallic densities. We want
to discuss this further.

We rewrite the self-energy as
\begin{eqnarray}\label{eq:18}
&&\Sigma^{xc}_{nn^{'}}({\bf k},\omega)=-{1\over 2\epsilon_0}
\sum_{{\bf k}^{'}n^{''}} U_{nn^{''},n^{'}n^{''}}({\bf k-k}^{'}) \nonumber \\
&&-{1\over 2\epsilon_0}
\sum_{{\bf k}^{'}n^{''}}^{{\rm occ}} {U_{nn^{''},n^{'}n^{''}}({\bf k-k}^{'})
(\omega-\varepsilon_{{\bf k}^{'}n^{''}}+\mu_0)
\over \omega-\varepsilon_{{\bf k}^{'}n^{''}}+\mu_0+\omega_p} \\
&&-{1\over 2\epsilon_0}
\sum_{{\bf k}^{'}n^{''}}^{{\rm unocc}} {U_{nn^{''},n^{'}n^{''}}({\bf k-k}^{'})
(\omega-\varepsilon_{{\bf k}^{'}n^{''}}+\mu_0)
\over \omega-\varepsilon_{{\bf k}^{'}n^{''}}+\mu_0-\omega_p} \nonumber
\end{eqnarray}
In the first term the sum is over the whole Brillouin zone, and it can
 therefore be shown that within the model in Eq.~(\ref{eq:11})
it is state independent. We can then focus on the next two terms. 
If $\omega_p$ is very large, these two terms go to zero.     The band 
is then just shifted to lower energies, without any change in shape 
or width. We next consider finite values values of $\omega_p$, 
but for a moment we assume that $\omega_p$ is still larger than the band 
width. If $\omega+\mu_0$ 
is put at the bottom of the band, both terms are then positive for all 
${\bf k^{''}}$. States at the bottom of the band are then pushed upwards
by the last two terms. 
In the same way, states at the top of the band are pushed downwards,
leading to a narrowing of the band. These arguments are not qualitatively
different if $\omega_p$ is somewhat smaller than the noninteracting band width. 
Some of the energy denominators in Eq.~(\ref{eq:18}) could become 
negative, but the effect of reducing $\omega_p$ tends to be a further        
reduction of the band width.                          
We observe that these argument are specific for this model, where we
have assumed that the states have not just a lower bound but also an upper 
bound. For the 
electron gas and most other systems there is no upper bound, and no 
definite statements of this type can be made.

In the calculation above, we have only considered the coupling to
the $t_{1u}$ charge carrier plasmon and neglected the coupling to,
e.g., the plasmon at about 25 eV as well as the exchange interaction
with all occupied states except the $t_{1u}$ states. These effects 
were considered in the calculation by Shirley and Louie,\cite{Louie}
who found a broadening of the $t_{1u}$ band by about 30 $\%$. If this
broadenig is added to our results we find an essentially unchanged
$t_{1u}$ band width. This result can then not explain 
the narrow Drude peak in the optical conductivity, but it is essentially
consistent with the rather small specific heat deduced for
these systems.

\end{multicols}

\begin{thebibliography}{*}

\bibitem{Sohmen}E. Sohmen, J. Fink, and W. Kr\"atschmer, Europhys. Lett.
{\bf 17}, 51 (1992).

\bibitem{Eyert}
O. Gunnarsson, V. Eyert, M. Knupfer, J. Fink, J.F. Armbruster, 
J. Phys.: Condens. Matter {\bf 8}, 2557 (1996).

\bibitem{Liechtenstein}
A.I. Liechtenstein, O. Gunnarsson, M. Knupfer, J. Fink, and J.F. Armbruster,
J. Phys.: Condens. Matter {\bf 8}, 4001 (1996).

\bibitem{Knupfer}
M. Knupfer, M. Merkel, M.S. Golden, J. Fink, O. Gunnarsson, V.P.
Antropov, Phys. Rev. B {\bf 47}, 13944 (1993).

\bibitem{Hedin}
L. Hedin, Phys. Rev. {\bf 139}, A796 (1965).

\bibitem{DeGiorgi}
L. Degiorgi, Mod. Phys. Lett. B {\bf 9}, 445 (1995).

\bibitem{Brink}
J. v. d. Brink, O. Gunnarsson, and V. Eyert, (to be published).


\bibitem{Meingast}
G.J. Burkhart, and C. Meingast, Phys. Rev. B {\bf 54}, R6865 (1996). 

\bibitem{RevModPhys}
O. Gunnarsson, Rev. Mod. Phys. (in press).
 
\bibitem{Lundqvist}
B.I. Lundqvist, Phys. kondens. Materie {\bf 6}, 193 (1967); 
{\it ibid} {\bf 6}, 206 (1967); L. Hedin, B.I. Lundqvist,
and S. Lundqvist, Int. J. Quantum Chem. {\bf IS}, 791 (1967).

\bibitem{Overhauser}
A.W. Overhauser, Phys. Rev. B {\bf 3}, 1888 (1971).

\bibitem{Louie}
E.L.~Shirley and S.G.~Louie, Phys.~Rev.~Lett.~{\bf 71}, 133 (1993).

\bibitem{Langreth} D. Langreth, Phys. Rev. B {\bf 1}, 471 (1970).

\bibitem{HedinPS} L. Hedin, Physica Scripta {\bf 21}, 477 (1980).

\bibitem{Migdal}
O. Gunnarsson, V. Meden, and K. Sch\"onhammer, Phys. Rev. B
{\bf 50}, 10462 (1994).
 
\bibitem{Sawatzky}
R.W. Lof,  M.A. van Veenendaal, B. Koopmans, H.T. Jonkman, and G.A.
Sawatzky, 1992, Phys. Rev. Lett. {\bf 68}, 3924.

\bibitem{C60Mott}O. Gunnarsson, E. Koch, and R.M. Martin, Phys. Rev. B
{\bf 54}, R11026 (1996).

\bibitem{eps}A.F. Hebard, R.C. Haddon, R.M. Fleming, and R. Kortan,
Appl. Phys. Lett. {\bf 59}, 2109 (1991).

\bibitem{Orientation}O. Gunnarsson, S. Satpathy, O. Jepsen, and O.K. Andersen,
Phys. Rev. Lett. {\bf 67}, 3002 (1991).

\bibitem{Satpathy}
S. Satpathy,  V.P. Antropov, O.K. Andersen, O. Jepsen, O. Gunnarsson,
and A.I. Liechtenstein, Phys. Rev. B {\bf 46}, 1773 (1992).


\end{thebibliography}
\end{document}